\documentclass[aps,prfluids, reprint, superscriptaddress]{revtex4-2}

\usepackage{import}
\usepackage{lipsum}

\usepackage{natbib}
\usepackage{times}
\usepackage{graphicx}
\graphicspath{{fig/}}
\usepackage{amsfonts}
\usepackage{amsmath,amsthm,amssymb,mathrsfs}
\usepackage{dsfont}
\usepackage{xcolor}
\usepackage{microtype}
\usepackage{bm}
\usepackage{siunitx}
\usepackage[colorlinks=true, allcolors=black, citecolor=blue, linkcolor=blue, urlcolor=blue]{hyperref}
\usepackage{verbatim}
\usepackage{diagbox}

\usepackage{float} 
\usepackage{subfigure}
\usepackage{booktabs}
\usepackage{soul}

\usepackage{algpseudocode}

\usepackage{xcolor}       
\usepackage{listings}     

\lstdefinelanguage{Matlab}{
    morekeywords={break, case, catch, continue, else, elseif, end, for, function, global, if,
        otherwise, persistent, return, switch, try, while},
    sensitive=true,
    morecomment=[l]\%,         
    morestring=[m]',           
    morestring=[m]"            
}

\usepackage[capitalize]{cleveref}

\renewcommand{\Im}[1]{\mathrm{Im}(#1)}

\begin{document}

\title{Rediscovering shallow water equations from experimental data}

\author{Kjell S. Heinrich}
\thanks{K.S.H. and D.S.S. are to be considered joint first authors.}
\affiliation{Department of Energy and Process Engineering, Norwegian University of Science and Technology, 7491 Trondheim, Norway}
\author{Douglas S. Seth}
\thanks{K.S.H. and D.S.S. are to be considered joint first authors.}
\affiliation{Department of Mathematical Sciences, Norwegian University of Science and Technology, 7491 Trondheim, Norway}
\author{Mats Ehrnström}
\affiliation{Department of Mathematical Sciences, Norwegian University of Science and Technology, 7491 Trondheim, Norway}
\author{Simen Å. Ellingsen}
\affiliation{Department of Energy and Process Engineering, Norwegian University of Science and Technology, 7491 Trondheim, Norway}

\begin{abstract}
    New data-driven methods have advanced the discovery of governing equations from observations, enabling parsimonious models for complex systems. Here, we 'rediscover' a shallow-water equation closely related to Korteweg--de Vries (KdV) using only video recordings of solitons in a simple flume. Two fundamentally different approaches --- weak-form sparse identification of nonlinear dynamics (WSINDy) and a novel Fourier-multiplier method --- recover the same PDE, demonstrating that the equation is inherent in the data and robust to the choice of method. Both identify the same terms with comparable magnitudes and errors. To validate the models, we solve the discovered equations forward in time and compare them with additional experimental cases that were not used in the discovery. Based on the results, we discuss absolute and cumulative errors, as well as the strengths and limitations of the two discovery approaches. Together, these results demonstrate the potential of equation discovery from everyday experiments ('GoPro physics') and highlight shallow-water waves as an ideal test bed for developing and benchmarking new methods.
\end{abstract}

\maketitle
\section{Introduction}
\label{sec:intro}

Nearly every branch of natural science and engineering depends on the predictive ability and interpretable power of partial differential equations (PDEs), yet the systems under study are frequently complex, including emergent phenomena whose dynamics are not known or understood. Discovery of nonlinear differential equations directly from measured data is, therefore, a highly attractive prospect, and much progress has been made in recent years (see, e.g., \citet{brunton24} for a recent review). If a PDE can be found that is able to predict future observations and explain past ones, one can eschew the use of expensive, inscrutable `black-box' deep-learning methods. Instead, dynamic evolution equations can be expressed in terms of concepts that can be understood in physical terms. This would allow for a new and deeper understanding of the vast body of work on classical dynamical system equations derived from first principles in the natural and technical sciences. Additionally, evolution equations discovered from data also combine effectively with machine learning methods for prediction and data assimilation \cite{cheng2023,gao2025}.

A natural first step in developing and building faith in equation discovery methods is to attempt to `rediscover' well known equations from experimental or numerical data of processes where the underlying physics is understood and where there exist simple, interpretable equations which are known to describe the system. Typically such equations are derived for asymptotic cases and found to provide good approximations also when the underlying conditions are not strictly satisfied. Our approach here is to study a physical system using observations made with common, everyday equipment --- nicknamed 'GoPro physics' \cite{makke2024} --- for a phenomenon which in similar circumstances has been shown to be effectively captured by a known equation, derivable from first principles. 

Our goal in this work is two-fold: 
\begin{enumerate}
    \item To demonstrate how the governing equation of a classical system can be found from a benchtop experiment with everyday equipment, and that two very different methods give the same result.
    \item To present, test and validate a new equation-discovery method based on Fourier multipliers and compare its performance to another, established and fundamentally different method.
\end{enumerate}

We have chosen waves in shallow water as our physical system, a suitable choice for several reasons. It is a classical system, studied for a century and a half \cite{boussinesq1877,korteweg1895}, the motion of the water surface in time and space is approximated in theory by simple and much-studied PDEs, and a deliberately crude experiment can be realised easily, quickly and at low cost and no special equipment. 

Key to our approach, however, is that we assume very little \emph{a priori} knowledge of the form of an underlying equation but let the data speak almost entirely for itself. Since we wish to find equations that are interpretable and as simple and parsimonious as possible, the restrictions we must by necessity place on what types of equations to allow are no limitation in practice. Our goal is not to study shallow-water waves in our makeshift lab in themselves or to show that they adhere to a specific theory. On the contrary we demonstrate how the fact that the `discovered' equation is \emph{not} exactly the theoretical expectation is fertile grounds for identifying interesting processes which the idealised theory does not capture. As preparation for a systematic investigation, an equation discovered by rough means can guide the researcher where to look. Data-driven equation discovery thus opens up potential avenues for revealing new aspects even for well-known, classical systems \cite{gao2025}. 

\subsection{Shallow-water waves; the KdV equation}

One of the best-known nonlinear differential equations in physics is the Korteweg--de Vries (KdV) equation \cite{korteweg1895}, a partial differential equation in time and one spatial dimension which describes nonlinear surface waves propagating in shallow water \citep[e.g.,][\S 31]{wehausen1960}. The KdV equation has been extensively validated against careful experiments going back many decades (e.g., \cite{zabusky1971}). Assuming $Y(\chi,\tau)$ is a suitably nondimensional surface elevation (the local vertical position of a water surface) and $\chi$ and $\tau$ are likewise the nondimensional spatial coordinate and time, the equation has the canonical form
\begin{equation}\label{eq:KdV}
    \partial_\tau Y = 6Y \partial_\chi Y -\partial_\chi^3 Y = 3\partial_\chi Y^2 -\partial_\chi^3 Y.
\end{equation}
Various generalisations exist \cite{horikis2022extended}, and higher-order derivatives may be present in the setting of critical Bond number or by expanding to higher order in the shallowness parameter, such as in the fifth-order Kawahara equation \cite{kawahara1972oscillatory}.
 
As a test bed, shallow-water waves are an attractive system to study due to the relative ease of obtaining experimental data with simple means and the availability of transparent analytical theory for comparison. We make shallow-water soliton waves in a small, makeshift wave flume using a manual, hand-held piston as wave-maker, and film the moving water surface with an uncalibrated camera. The presence of uncontrolled noise sources, variation between trials, lack of calibration, and no detailed control over wave generation whatsoever --- beyond making waves that look soliton-like to the naked eye --- allows us to test the robustness of our equation discovery procedures.

Two independent methods are employed to extract a parsimonious partial differential equation of motion from the videos. Although the physical system is expected to be approximately (and only approximately) described by the KdV equation, our methods only assume that the change in time can be expressed by spatial derivatives.

The factor $6$ in Eq.~\eqref{eq:KdV} (and consequently $3$ in the last form) is conventional, and convenient for some mathematical properties, but in fact depends on the chosen scaling of time and space. Here we will use characteristic time and length scales constructed from the mean depth $h$ and gravitational acceleration $g$, as is the common choice. Let now the surface elevation, spatial and temporal coordinates in dimensional units be $\eta^*$, $x^*$ and $t^*$, and define nondimensional quantities $\eta, x$ and $t$ according to
\begin{equation}
    t^* = \sqrt{h/g}\, t, ~~ x^* = h\, x, ~~ \eta^* = h\, \eta.
\end{equation}
The normalized nondimensional equation \cite{dingemans1997water} is
\begin{align}
    \displaystyle \partial_t \eta + \partial_x \eta + \frac{3}{2} \eta\, \partial_x \eta + \frac{1}{6} \partial_x^3 \eta = 0.
\end{align}
 
In \cref{sec:experiment} we present the experimental setup and edge detection to extract the surface elevation from the videos. In \cref{sec:methods} we introduce the two methods used to identify the governing equations. The first is the 'Sparse Identification of Nonlinear Dynamics' (SINDy), which was first introduced by \citet{brunton2016discovering}. We use its weak-form formulation \cite{reinbold2020using,messenger2021weakODE,messenger2021weakPDE}, which we explain in \cref{subsec:SINDy}. In \cref{subsec:Fourier} we introduce the second method, a novel method for equation discovery in Fourier space. We present the identified equations in \cref{sec:results}, along with simulations that we compare with solitons withheld in the discovery. Finally, \cref{sec:discussion} discusses the results and differences in the methods. 

\section{Experiment and preprocessing}\label{sec:experiment}

The data was collected using a simple experimental setup consisting of a wave flume measuring $261\times46 \text{ cm}^2$, with a water depth of 32 mm. Solitons were generated by manually pushing a piston that matched the width of the flume. At this depth, the Bond number is small $\mathcal{B} = \gamma / (\rho g h^2) \approx 7\times 10^{-3}$ where $\gamma\approx 0.073$\,N/m is the water-air surface tension coefficient, so the dynamics are gravity-dominated. This setup was intentionally designed to avoid additional control over the parameters. We embraced this lack of control to test the equation-identification procedure under realistic benchtop conditions, including hand-made waves, imperfect boundary conditions, and consumer-level imaging. Our goal is to ensure that reproducibility does not depend on hardware or calibration. Instead, we aim to test that the methods consistently identify a leading-order balance, even when faced with small, uncontrolled variations between trials. The camera captured a length of $60$\,cm of the water surface in the flume, showing fluorescence induced by a laser pointer from fluorescein in the water. While this illumination improves contrast, it is not essential. The analysis depends on surface elevation, which can be accurately extracted from videos recorded with, say, a mobile phone camera using a standard edge-detection algorithm \cite{cao2024identification}.

\begin{figure}[hbt]
    \centering
    \includegraphics[width=0.5\textwidth]{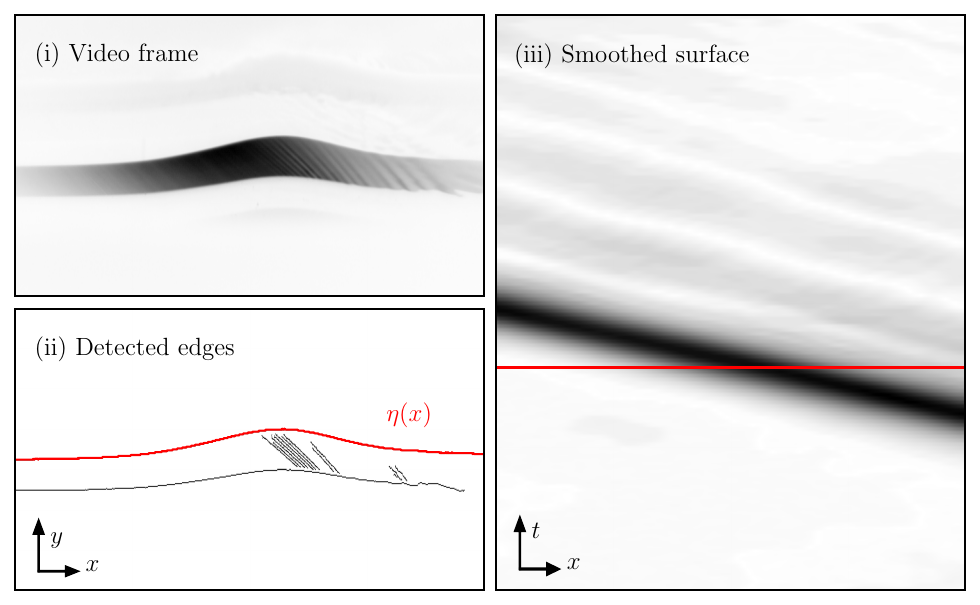}
    \caption{(i) Example frame from recorded video, (ii) illustration of the edge detection algorithm, (iii) full $(x,t)$ field.}
    \label{fig:placeholder}
\end{figure}

The videos captured during the experiment show a side view of the solitons. To extract the free surface, we began by identifying all edges in a single frame using the Canny algorithm, which is designed to detect the most significant gradients \cite{canny1986computational}. We kept only the uppermost edge, which was interpreted as the surface in each frame. In the final step, the surface $\eta$ was smoothed using the method described in \citet{cao2024identification}. This procedure was applied to all frames across all videos. \Cref{fig:placeholder} illustrates qualitatively the process for one soliton, while a detailed explanation is given in the appendix.

\section{Methods}\label{sec:methods}
After the processing, we have $N$ non-dimensional fields ${\eta_k(x, t) \in \mathbb{R}^{l\times m_k}}$ of the surface elevation, where ${k\in(1,N)}$. All videos are cropped the same in the $x$-direction, so $l$ denotes the spatial dimension, and $m_k$ the temporal dimension for soliton $k$. These fields will be combined into a single data matrix $H$. We add them in succession, such that 
\begin{align} \label{eq:def_H}
    \displaystyle H(x, t) = \big[\eta_1(x,t)\:\: \eta_2(x,t) \:\:\dots\:\: \eta_{N}(x,t) \big]\:,
\end{align}
where the matrix has the dimensions ${H\in \mathbb{R}^{l\times\sum_k m_k}}$. The goal is to find an equation that can describe the dynamics of the solitons captured on film. With both methods, we choose to identify an equation on the form
\begin{align} \label{eq:zeroth_step}
    \displaystyle \partial_t H = f(H) .
\end{align}
We analyze all videos simultaneously to minimize the likelihood of identifying a poor equation, one that does not accurately describe the observed motion. There is not a single, unique equation that fully explains the set of states or dynamics in our setup. Since the data contains only a finite number of points, an infinite number of equations can describe the dynamics when they are allowed to be arbitrarily complex. Therefore, our goal is to find the simplest possible description that still captures the fundamental characteristics of the waves. We aim to find the solution that strikes a balance between error and complexity. By examining all videos together, we can identify the equation that governs the behavior across all the videos. To meassure how well our equations are, we define the real-space error given by
\begin{align} \label{eq:error_real_space}
    \displaystyle E^\textrm{real} = \frac{\|\partial_t H - f(H)\|_F}{\|\partial_t H\|_F},
\end{align}
where $\|A\|_F \equiv \sqrt{\text{Tr}(A A^T)}$ is the Frobenius Norm. 

In the hope of obtaining a result which is readily interpretable, we search for equations where the right-hand side of \eqref{eq:zeroth_step} depends only on $H$ and its spatial derivatives, i.e.,
\begin{align} \label{eq:Starting_point}
    \displaystyle f(H) = \sum_{q,p} c_{q,p}\: \partial^q_x\big(H^p\big),
\end{align}
where $\partial_x^q$ denotes the $q$-th order spatial derivative and $H^p$ the $p$-th power of $H$. Note that by 'power', we mean the $p$-th-order Hadamard product of $H$, and \textit{not} the ordinary matrix multiplication. The powers are applied elementwise, not to the matrix itself. Since $H$ contains some noise, the derivatives in \cref{eq:error_real_space} are estimated by alternating between a first-order finite difference and smoothing with a Savitzky--Golay filter for the higher-order derivatives.

With \cref{eq:zeroth_step,eq:Starting_point}, we make two assumptions in the equation discovery that hold true for both methods:
\begin{enumerate}
    \item We can find a simple, interpretable equation that describes the dynamics via $\partial_t H$.
    \item We can describe $\partial_t H$ with only spatial derivatives $\partial_x^q$ applied to $H$ and $H^2$. 
\end{enumerate}
The methods differ in terms of \textit{at what stage} assumption two is applied, and the order of $q$. While WSINDy restricts the allowed values of $q$ from the start, the Fourier multiplier method does so as a last step to enable comparison between the methods.

\subsection{Weak-form SINDy and Sparse Regression}\label{subsec:SINDy}

The goal of the SINDy algorithm is to find the optimal coefficients in \cref{eq:Starting_point}. While there are many extensions of the algorithm \cite{rudy2017data,kaheman2020sindy,fasel2021sindy,fasel2022ensemble,zolman2024sindy,zheng2024sindy,viknesh2024adam}, our focus is on the variant that does not rely on estimating spatial derivatives in the discovery, since noise tends to be amplified by differentiation. We introduce a unimodal polynomial test function $\psi_k(x,t)$ with enforced compact support on ${\Omega_k = \{(x,t) : |x - x_k| \leq N_x,; |t - t_k| \leq N_t\}}$. The position $(x_k,t_k)$ is the center of the window and $(N_x,N_t)$ are the corresponding half-widths in space and time \cite{messenger2021weakODE,messenger2021weakPDE}. Multiplying \cref{eq:zeroth_step} by $\psi_k(x,t)$ and integrating over $\Omega_k$, we get
\begin{align} \label{eq:weak_SINDY_problem}
    \displaystyle X^{(k)} = \sum_{p,q} c_{q,p}\: \theta_{q,p}^{(k)}\:,
\end{align}
where we have defined 
\begin{align}
    \displaystyle X^{(k)} \equiv& -\int_{\Omega_k} H(x,t)\: \partial_t \psi_k\: \mathrm{d}\Omega,\label{eq:integration_definitions1}\\
    \displaystyle \theta_{p, q}^{(k)} \equiv&\, (-1)^q\int_{\Omega_k}  H^p(x,t)\: \partial_x^q \psi_k\: \mathrm{d}\Omega .\label{eq:integration_definitions2}
\end{align}
The unknowns in \cref{eq:weak_SINDY_problem} are $c_{q,p}$, which are independent of the chosen domain $\Omega_k$. Thus, we integrate on $K$ randomly selected domains and organize the results into column vectors. We use a library containing the terms ${\theta = \{H,\: \partial_x H,\:  \dots,\: \partial_x^5H\:, H^2,\: \partial_x H^2 \}}$ from the experimental data, fixing the orders of differentiation, $q$. It results in the following matrix equation,
\begin{align} \label{eq:matrix_eq}
    \hspace*{-1mm}\displaystyle \begin{pmatrix}
        X^{(1)} \\ X^{(2)} \\ \vdots \\ X^{(K)}
    \end{pmatrix} = \begin{pmatrix}
        \theta^{(1)}_{0,1} & \theta^{(1)}_{1,1} & \dots & \theta^{(1)}_{5,1} & \theta^{(1)}_{1,2}\\
        \theta^{(2)}_{0,1} & \theta^{(2)}_{1,1} & \dots & \theta^{(2)}_{5,1} & \theta^{(2)}_{1,2}\\
        \vdots & \vdots & \ddots & \vdots & \vdots\\
        \theta^{(K)}_{0,1} & \theta^{(K)}_{1,1} & \dots & \theta^{(K)}_{5,1} & \theta^{(K)}_{1,2}
    \end{pmatrix} \begin{pmatrix}
        c_{0,1}\: \\ c_{1,1}\: \\ \vdots\\c_{5,1}\: \\ c_{1,2}\:
    \end{pmatrix}
\end{align}
which can be expressed as $\bm{X} = \Theta \bm{c}$. Here, $\bm{c}$ represents the vector of unknown coefficients. The above approach is known as weak-SINDy (WSINDy), as differentiation operators are shifted to analytical test-functions through integration by parts \cite{reinbold2020using,messenger2021weakODE}. Numerically, these were implemented using the \verb|PySINDy| package in Python \cite{desilva2020, Kaptanoglu2022}.

When $K$ is larger than the size of the library and the number of unknown coefficients, \cref{eq:matrix_eq} is overdetermined. The problem can be solved using a least-squares regression. This approach, however, typically yields a coefficient vector $\bm{c}$ in which all entries are non-zero \cite{brunton2022data}. This suggests that all terms in the library are necessary to predict $\partial_t H$.  Our goal is to identify a sparser PDE that describes the dynamics, which means a sparser coefficient vector  $\bm{c}$. To avoid overfitting, we add two additional terms to the least-squares loss function
\begin{align} \label{eq:ElNet_loss}
    \hspace*{-1mm}\displaystyle \mathcal{L}(\bm{c}) = \frac{1}{2K} \| \bm{X} - \Theta \bm{c}\|^2_2 + \alpha \sigma \|\bm{c}\|_1 + \frac{\alpha}{2} (1-\sigma)\|\bm{c}\|_2^2,
\end{align}
where $\alpha$ is a sparsity promoting knob and ${0 \leq \sigma \leq 1}$ fixes the ratio of the $L_1$ and $L_2$ norm. The $L_{1}$ penalty promotes sparsity by driving many entries of $\bm{c}$ exactly to zero. The $L_{2}$ penalty shrinks coefficients toward zero without forcing exact zeros, improving numerical stability under collinearity and reducing overfitting. \Cref{eq:ElNet_loss} is called the Elastic net \cite{zou2005regularization}, and for $\sigma=1$ it reduces to LASSO \cite{tibshirani1996regression}. We have implemented a 10-fold cross-validated Elastic Net regression with sequential thresholding to ensure sparsity of $\bm{c}$. In this optimization problem, we find the optimal $\alpha$ and $\sigma$ for the Elastic Net using cross-validation with \verb|scikit-learn| in Python \cite{scikit-learn}. The implementation algorithm is given in the appendix. 

To confidently identify relevant factors in the governing PDE, we perform the WSINDy regression $M$ times. Since the integration domains are chosen randomly, we get a statistical distribution of coefficients $\bm{c}_m$. These coefficients provide inclusion probabilities and distributions that we can use to uncover the dynamics in the data. The size of $M$ is chosen to stabilize the ensemble statistics, ensuring that the chosen terms and estimated coefficients vary little with additional runs. In practice, $M = 100$ for exploratory analyses and $M = 1000$ for final estimates. Additionally, we analyze the mean error of the WSINDy regression, defining it as 
\begin{align} \label{eq:SINDy_regression_error}
    \displaystyle E^\textrm{reg} = \frac{1}{M \sqrt{K}} \sum_{m=1}^M \frac{\|\bm{X}_m - \Theta_m \bm{c}_m\|_2}{\|\bm{X}_m\|_2} ,
\end{align}
which provides a scale-invariant measure of prediction accuracy across a varying number of weak integration domains.

\subsection{Fourier multipliers} \label{subsec:Fourier}

The second method used to construct a governing equation from the data relies on using partly overlapping, and partly complementary, assumptions on the form of the equation. Just as in the first method, this form covers a wide class of model equations for the water wave problem \cite{Lannes_2013}, including the KdV equation. As in \cref{subsec:SINDy}, we assume that the governing equation can be written as
\begin{align}\label{eq:FourierMultMethod}
    \partial_t\eta=L\eta+N\eta^2,
\end{align}
where in this case $L$ and $N$ are general Fourier multipliers, so that
\begin{align}
\widehat{L\eta}(\xi, t)=\ell(\xi)\widehat{\eta}(\xi, t)
\end{align}
and
\begin{align}
\widehat{N\eta^2}(\xi, t)=n(\xi)\widehat{\eta^2}(\xi, t),
\end{align}
where the hat denotes the spatial Fourier transform ${\widehat{f}(\xi, t)=\tfrac{1}{\sqrt{2\pi}}\int_{-\infty}^{\infty}f(x, t)e^{-i x\xi}\,\mathrm{d}x}$. The functions $\ell$ and $n$ are called the symbols of the operators $L$ and $N$, respectively. Note that these operators are differential operators exactly when the symbol is a polynomial, so these are more general operators than the ones in \cref{eq:Starting_point}. However, we will assume the symbols to be odd to match the dispersive character of the shallow-water waves. In the special case of differential operators, this is equivalent to excluding all even orders of derivatives. {This constitutes a restriction on $q$ in \cref{eq:Starting_point}, but will first be applied after we have determined } the symbols $\ell$ and $n$ by estimating $\partial_t \widehat{\eta}(\xi, t)$, $\widehat{\eta}(\xi, t)$, and $\widehat{\eta^2}(\xi, t)$ from the data, $H$ (introduced in \cref{subsec:SINDy}), as described below. 

We write $\widehat{H}(\xi_i)$ for the spatially discrete Fourier transform (DFT), here obtained through the fast Fourier transform, at angular frequency $\xi_i$ (we use angular frequency so that polynomial coefficients in frequency space directly match the coefficients $c_{q,p}$ for a differential operator in real space); and let $\widehat{H^2}(\xi_i)$ be the corresponding DFT of $H^2$, where the square is taken element-wise in the matrix. To approximate the time derivative of $\widehat{H}$, we concantinate the derivatives of each individual field, ${\partial_t \widehat{H} = [\partial_t\hat{\eta}_1 \,\, \partial_t\hat{\eta}_2\, \dots]}$, similar to \cref{eq:def_H}.  We approximate the time derivative of each soliton as
\begin{equation} \label{eq:Fourier_time_derivative}
    \partial_t\hat{\eta}_k(\xi_i,t_j)=\mathcal{F}^{-1}_t[ i\omega \mathcal{F}_t[\hat{\eta}_k(\xi_i,\cdot)](\omega)](t_j),
\end{equation}
where $\mathcal{F}_t$ here denotes the temporal DFT transforming to the angular frequency $\omega$. Similar results can be obtained with a finite difference method, but a high order method is required due to relatively large time steps in the data. We obtain $\sum_{k=1}^N m_k$ linear equations
\begin{align}
\partial_t \widehat{H}(\xi_i)=\ell(\xi_i)\widehat{H}(\xi_i)+n(\xi_i)\widehat{H^2}(\xi_i)
\end{align}
for $\ell(\xi_i)$ and $n(\xi_i)$. Equivalently, this can be written as ${Y_i=X_iA_i}$ with
\begin{align}
A_i=\begin{pmatrix}
        \widehat{H}(\xi_i)\\
     \widehat{H^2}(\xi_i)
\end{pmatrix},\qquad Y_i=\begin{pmatrix}\partial_t\widehat{H}(\xi_i)
\end{pmatrix},
\end{align}
and
\begin{align}
X_i=\begin{pmatrix}
    \ell(\xi_i)& n(\xi_i)
\end{pmatrix}
\end{align}
Taking $X_i$ as the least-squares minimizer
\begin{equation}\label{eq:least-squares minimizer}
X_i=Y_iA_i^{\dagger}(A_i A_i^{\dagger})^{-1},
\end{equation}
for these equations gives us $\ell(\xi)$ and $n(\xi)$ at the $l$ sampled frequencies $\xi_i$ of the discrete Fourier transform. Here $A^\dagger_i$ is the conjugate transpose of $A_i$. In practice, we do not use all time steps and frequencies. This restriction is described in detail in \cref{subsec:results_FT}. 

The symbols $\ell$ and $n$ can be taken as they are for a discrete spectral equation, or inter- or extrapolated if one wants global or continuous expressions. In this article, we use polynomial interpolation yielding a partial differential equation to facilitate comparison with the WSINDy method.

\renewcommand\arraystretch{1.5}
\begin{table*}
    \centering
    \begin{tabular}{lcccccccccccc}
        \toprule
         & $\quad{\ }$ & $\quad H\quad$ & $\quad\partial_x H\quad$ & $\quad\partial_x^2 H\quad$ & $\quad\partial_x^3 H\quad$ & $\quad\partial_x^4 H\quad$ & $\quad\partial_x^5 H\quad$ & $\quad H^2\quad$ & $\quad\partial_x H^2\quad$ & $\quad{\ }$ & $\quad E^\textrm{real}\quad$ & $\quad\overline{E^\textrm{cum}(T)}\quad$ \\
        \hline
        WSINDy $\:$&& - & 0.804 & - & 0.616 & - & 0.059 & - & 1.362 && 29.71\% & 6.27 \%\\
        Fourier multipliers $\:$&& - & 0.848 & - & 0.516 & - & 0.059 & - & 1.367 && 30.14\% & 4.02 \%\\
        \bottomrule
    \end{tabular}
    \caption{Identified coefficients for the two different methods and the error in real space. The last column gives the mean of the total cumulative average error of models built by the coefficients, solved forward in time, and compared to the experimental data of solitons that were not used in the equation discovery.}
    \label{tab:Results_summary}
\end{table*}

\newpage
\section{Results} \label{sec:results}

Using both identification routes, weak-form sparse regression and the Fourier-multiplier fit, we recover the same KdV-type evolution law
\begin{align} \label{eq:identified_equation}
    \hspace*{-1mm}\partial_t H = c_{1, 1}\: \partial_x H + c_{3, 1}\: \partial_x^3H + c_{5,1}\: \partial_x^5 H + c_{1,2}\: \partial_x H^2 .
\end{align}
The corresponding coefficients are listed in \cref{tab:Results_summary}. With these parameters fixed from training, forward integration of \eqref{eq:identified_equation} reproduces the shape and speed of solitons withheld from the regression, without any post-hoc tuning. We first present the regression results of WSINDy in \cref{subsec:results_SINDy}, fixing the integration domain size $N_x \times N_t$ and $K$. Following that, we show the polynomial fit of the Fourier-multiplier operators in \cref{subsec:results_FT}. Finally, in \cref{subsec:results_forward}, we validate the model by simulating \cref{eq:identified_equation} forward in time and comparing the results against held-out test solitons. 

\subsection{WSINDy} \label{subsec:results_SINDy}

\begin{figure}[hbt]
    \centering
    \includegraphics[width=0.5\textwidth]{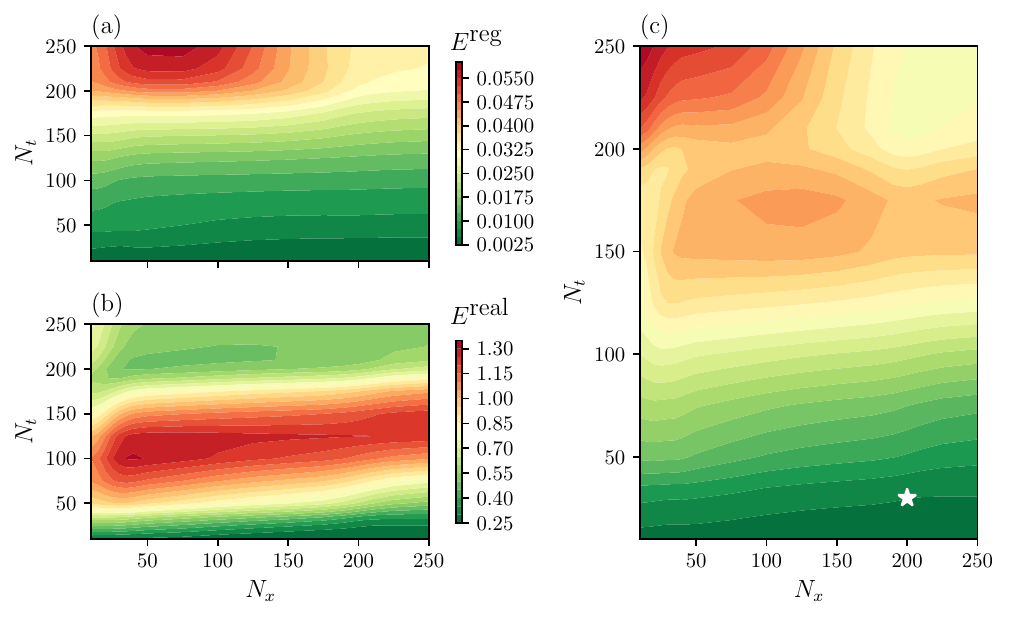}
        \caption{(a) Shows the regression error defined by \cref{eq:SINDy_regression_error} for ${M=100}$ and ${K=1000}$. (b) Shows the mean error in real space for the same coefficients as in (a). In (c), the errors are multiplied together, and the star shows the choice of domain size for the results after this.}
    \label{fig:Domainsize}
\end{figure}

The parameters chosen in the integration process, \cref{eq:integration_definitions1,eq:integration_definitions2}, significantly influence the identified equation. Therefore, we examine how  ${N_x \times N_t}$ and $K$ impact the regression error and the identified equation. Here, $N_x$ and $N_t$ represent the integration domain widths in the spatial and temporal dimensions, respectively, while $K$ denotes the number of randomly placed integration domains. In \cref{fig:Domainsize}, we show how the regression error given by \cref{eq:SINDy_regression_error}, varies with both $N_x$ and $N_t$. We also show the real-space error given by \cref{eq:error_real_space}. It is important to note that a good regression does not guarantee a reduced error in real space. In fact, varying the size of the domain can lead to significantly different equations. Therefore, even if the regression in weak space shows a low error, it is crucial to verify the resulting PDE in real space. If the right- and left-hand side of \cref{eq:Starting_point} differ significantly, it indicates that the equations may not accurately capture the correct dynamics. By analyzing the error in both spaces, we can find a suitable choice of domain size ${N_x \times N_t}$. The choice lands on ${N_x = 200}$ and ${N_t = 30}$ highlighted by a white star in \cref{fig:Domainsize}(c), which displays the multiplication of the errors in the two spaces. In \cref{fig:Domainsize}, the regression was done for $M=100$ because the coefficient statistics have stabilized well by that scale. Once $M$ exceeds a few dozen realizations, further increases yield diminishing changes in the mean errors and identified coefficients. We will defer the use of a larger $M$ for coefficient refinement to the next step.

\begin{figure}[hbt]
    \centering
    \includegraphics[width=0.5\textwidth]{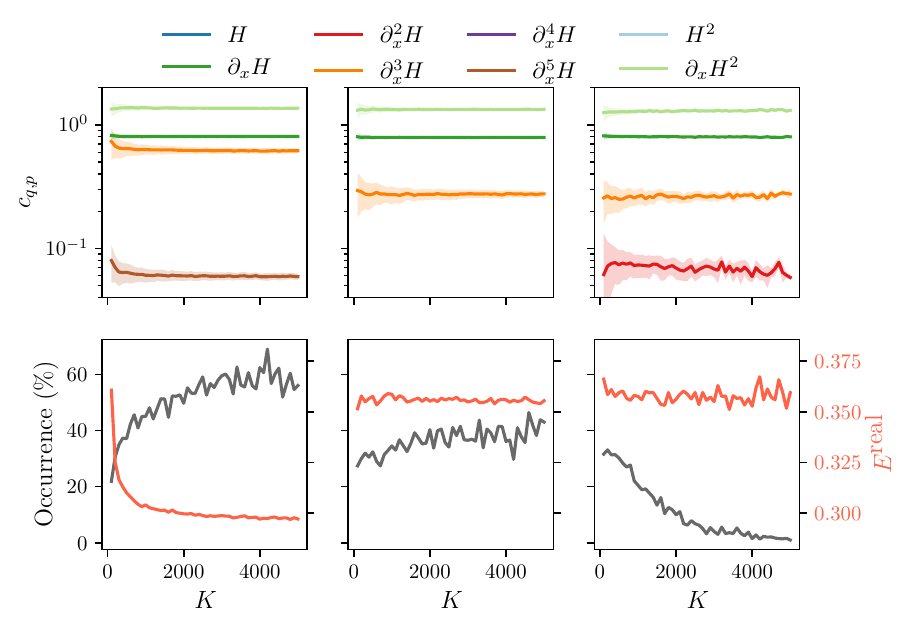}
    \caption{The upper row shows the conditional mean magnitude and variance of the coefficients $c_{\mu,\lambda}$ for $M=1000$ ensembles and ${N_x \times N_t = 200\times30}$ for different identified PDEs. The second row shows the real-space error and how often that PDE is identified.}
    \label{fig:Coefficients}
\end{figure}

For each ensemble, we get a slightly different coefficient vector $\bm{c}$ because the integration domains are sampled randomly. This provides a statistical distribution for each $K$, allowing us to calculate the mean of each entry in $\bm{c}$ as well as the probability of being identified. We observe that the magnitudes of the individual coefficients are influenced by each other. This is shown in \cref{fig:Coefficients}, where we present the conditional average and variance for the three most identified PDEs. For about $K \gtrsim 200$, these are the only three equations that are identified. Simply taking the mean of the coefficients from all $M$ ensembles would provide an inaccurate estimate of their magnitudes. For instance, the $\partial_x^3 H$ term, which is consistently identified across all regressions, would be assigned the incorrect value by a standard mean over all ensembles. \Cref{fig:Coefficients} shows that the number of times the first PDE, given in \cref{eq:identified_equation}, is identified dominates for large $K$. The figure also shows that the dissipation term with $\partial_x^2 H$ is present, but does not contribute significantly to the dynamics. The variance of all coefficients decreases along the $K$-axis, indicating that they are converging. We have chosen to perform the regression up to $K=5000$ with $M=1000$ ensembles. This choice ensures that the PDE in the rightmost column shown in \cref{fig:Coefficients}, which appears infrequently for large $K$, is sampled sufficiently to yield a reliable conditional mean and variance.

We include the fifth-order spatial derivative $\partial_{x}^{5} H$ based on the results shown in \cref{fig:Coefficients}. For large numbers of integration domains $K \gtrsim 4000$, the most frequently identified model, recovered in roughly $60\%$ of the ensembles, retains $\partial_x H,\: \partial_x^{3}H,\: \partial_x^{5}H$, and $\partial_x(H^{2})$ while the $\partial_x^{2}H$ term systematically vanishes. Although the competing PDEs are broadly comparable, the first one yields the lowest real-space error $E^\textrm{real}$. Moreover, the inferred coefficient of the fifth-order term is small in magnitude, $c_{5,1}\approx 0.06$, about an order below the leading coefficients, yet it remains consistently separated from zero. It stabilizes as $K$ increases, indicating that it is not a regression artifact but a weak, reproducible contribution. Considering the dominance of this structure across the ensemble, the systematic decrease in $E^\textrm{real}$, and the shrinkage of coefficient variability with $K$, these factors provide a clear justification for including $\partial_x^{5}H$ and excluding $\partial_x^{2}H$. The terms  $\partial_x H,\: \partial_x^{3}H$, and $\partial_x(H^{2})$ are identified for all regressions. 

\subsection{Fourier multipliers} \label{subsec:results_FT}

To obtain results for the Fourier multiplier method we need to make three significant choices: (i) pick which time steps of the data to use; (ii) determine which Fourier modes to include; (iii) choose the polynomials to approximate the Fourier multipliers. The last is necessary to turn the nonlocal \cref{eq:FourierMultMethod} into a PDE.

\begin{figure}[htb]
    \centering
    \includegraphics[width=0.5\textwidth]{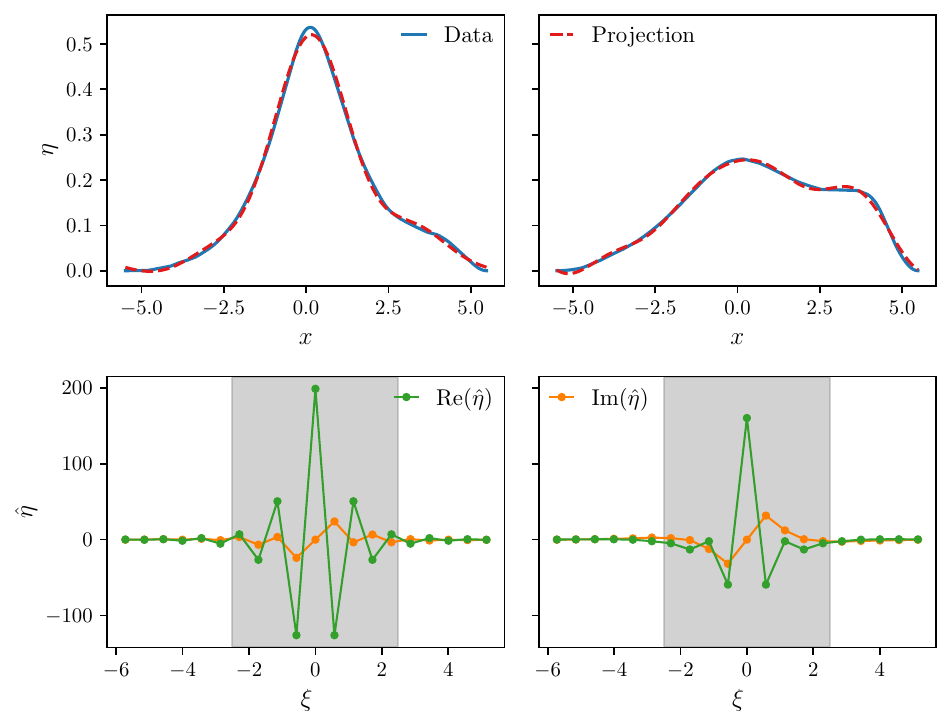}
    \caption{Top row: One high amplitude and one low amplitude soliton before (blue) and after (red) projection onto the used spatial frequencies.
    Bottom row: The real (green) and imaginary (orange) parts of the spectrum of the same solitons. The part of the spectrum that is used is shaded in gray.}
    \label{fig:Spectral_projection}
\end{figure}

We obtain the best results by only using a few time steps from when the wave is centered in the spatial domain. With this choice, we optimize with regard to the actual soliton and avoid the time steps when the wave enters or leaves the frame. These time steps give a distorted DFT due to the implicit periodization of the cut-off data. The soliton covers most of the $x$-axis, and even in frames with the soliton centered, we see small boundary effects --- the data and projection do not align perfectly at the edges --- due to the periodization. However, with a centered soliton, these effects are minimized (see \cref{fig:Spectral_projection}). For the analysis in this article, two time steps per soliton are sufficient.

We restrict the spatial frequencies because the high frequencies contain a significant amount of noise relative to the data. This noise comes from the experiment, data processing, and periodization. With more spatial data around the soliton, the periodization effects could be mitigated using a smooth window function, but such a function would significantly distort the shape of the actual soliton in our data. Instead, we find the least-squares minimizer for the zero mode and the four closest non-zero positive and negative modes. {This truncation effectively sets an upper limit for the highest order of spatial differentiation in the equation, $q$ in \cref{eq:Starting_point}.} These modes contain most of the relevant data (see \cref{fig:Spectral_projection}). We denote the indices for the chosen frequencies by $I$.

With these choices, we find $X_i$ as described in \cref{subsec:Fourier} for $\xi_i$ with $i\in I$, or, in short, $X=(X_i)_{i\in I}$. Computing the total relative error in the least-squares approximation, we obtain
\begin{align}
    E^{\mathrm{spectral}}(X)=\left(\sum_{i\in I}\frac{\Vert A_iX_i-Y_i\Vert_2^2}{\Vert Y_i\Vert_2^2}\right)^{1/2}=7.99\%.
\end{align}
However, as mentioned in \cref{subsec:Fourier} we take these multipliers to be odd. Since the data consists of real valued solitons, this is equivalent to only keeping the imaginary part of $X$. This could be imposed a priori, but doing it after minimization allows us to measure the increase in error this assumption entails. The error for only the imaginary part is $E^{\mathrm{spectral}}(i\Im{X})=11.92\%$. This can also be compared with the error from minimizing the multipliers with the assumption that they are imaginary, which yields $E^{\mathrm{spectral}}(X_\mathrm{Im})=11.34\%$.

The final step to obtain a PDE is to approximate the discrete multipliers with polynomials. Denote the polynomials $p_\mathrm{lin}^r(\xi)$ fitted to the imaginary part of $\ell$ and  $p_\mathrm{nonlin}^s(\xi)$ fitted to the imaginary part of $n$ of orders $r$ and $s$, respectively. Polynomials of order seven perfectly match the odd multipliers with the number of frequencies we used. The spectral resolution could be increased by zero-padding, but since we are seeking a simple model, we do not require higher-order polynomials.

\begin{figure}[hbt]
    \centering
    \includegraphics[width=0.5\textwidth]{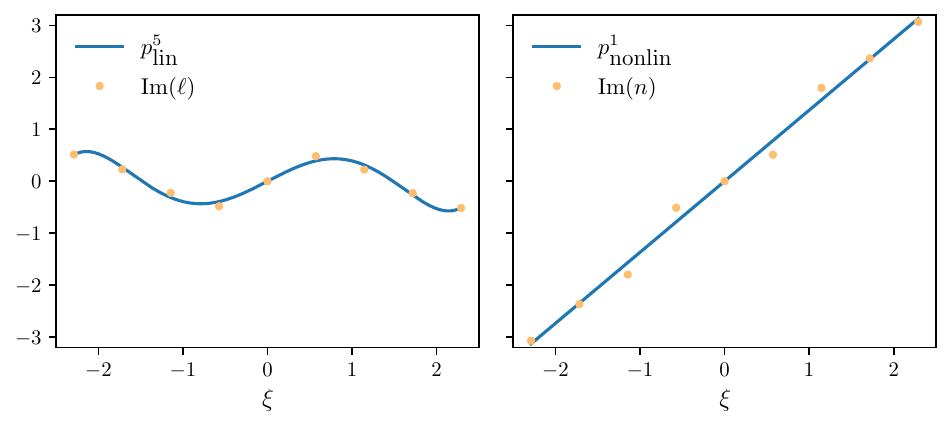}
    \caption{Left: The discrete linear multiplier (orange dots) and approximating polynomial $p^5_{\mathrm{lin}}$ (blue line).
    Right: The discrete nonlinear multiplier (orange dots) and approximating polynomial $p^1_{\mathrm{nonlin}}$ (blue line).}
    \label{fig:Multipliers}
\end{figure}

Setting
\begin{align}
    X^{r,s}=\begin{pmatrix}
    p_{\mathrm{lin}}^r(\xi_i)\\
    p_{\mathrm{nonlin}}^s(\xi_i)
\end{pmatrix}_{i\in I}
\end{align}
we can compute $E^{\mathrm{spectral}}(X^{r,s})$ for the various orders of approximating polynomials. The results are presented in \cref{tab:Polynomial_error}. Both low error and low-order polynomials are desirable. Therefore, we choose $r=5$ as the order for the linear polynomial and $s=1$ for the nonlinear polynomial. The multipliers obtained and the approximate polynomials chosen are shown in \cref{fig:Multipliers}. The resulting coefficients for the PDE are given in \cref{tab:Results_summary}.

\begin{table}
    \centering
    \begin{tabular}{c| c c c c}
    \toprule
        \diagbox{$s$}{$r$} & 1 & 3 & 5 & 7 \\
        \hline
      1 &48.72\% &23.79\% &12.35\% &17.38\% \\
       3 &47.23\% &22.61\% & 12.34\% &17.7\% \\
        5 &50.29\% &25.6\% & 12.08\% & 16.22\% \\
         7 &56.36\% &32.61\% &15.29\% &11.92\% \\
          \bottomrule
    \end{tabular}
    \caption{$E^{\mathrm{spectral}}(X^{r,s})$ for various orders of $p_{\mathrm{lin}}^r$ and $p_{\mathrm{nonlin}}^s$ approximating $\Im{\ell}$ and $\Im{n}$, respectively}
    \label{tab:Polynomial_error}
\end{table}

\subsection{Predictive skill} \label{subsec:results_forward}

In this subsection, we evaluate the predictive ability of the equations presented in \cref{tab:Results_summary}. We create simulations and compare them against solitons that were not included in the equation discovery, but were obtained from the same experimental setup. The PDE fitting detailed in \cref{subsec:results_SINDy} and \cref{subsec:results_FT} was conducted using videos of 18 solitons. For testing, we have a set of 7 solitons with nondimensional amplitudes ranging from approximately $A_k \approx 0.2$ to $A_k \approx 0.6$, covering the same amplitude range as the training solitons. 

The equations are solved with a finite difference scheme with a fourth order implicit Runge-Kutta method (SDIRK4 \cite{Hairer_1996}) for the time step and a second order central difference for the space variable. An implicit method is chosen both to ensure stability as well as being able to handle non-smooth initial data. We let the spatial discretizations be fixed by the experimental data and choose the time steps $\Delta t$ such that the accuracy in space and time are of similar size, that is $\mathcal{O}(\Delta t^4)\sim \mathcal{O}(\Delta x^2)$. Even with a fourth-order method in time, this requires about $10$ intermediate steps between each data frame. We impose a Dirichlet boundary condition on the side where the wave enters, given by a cubic spline of the data. At the other boundary, we extend the domain to allow the wave to exit the frame freely. The data was also aligned in time so that all waves entered at the same time for better comparison. To handle high frequency oscillations we apply Kreiss-Oliger dissipation \cite{KreissOliger1973}, adding a dissipative term $\epsilon\partial_x^6 \eta$ to the equations with $\epsilon=0.003$.

\begin{figure}[htb]
    \centering
    \includegraphics[width=0.5\textwidth]{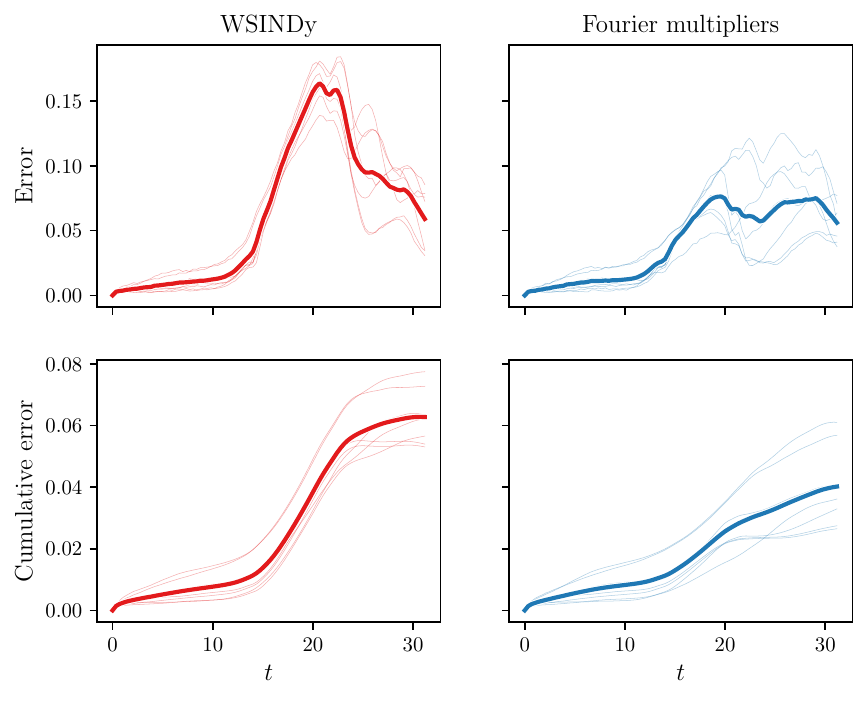}
    \caption{The absolute and cumulative average error for every model for the solitons in the test set. The mean errors at each time step are shown as a thicker line.}
    \label{fig:Error_modeling}
\end{figure}

To evaluate performance, we calculate the mean error per spatial index, normalized by the soliton amplitude, for each time step. This metric ensures that a larger sampling window does not yield a higher error, and it does not become ill-defined for time steps where the solitons have not entered the frame. Normalizing by the $L_2$-norm of $\eta_j$ instead would introduce such problems, since $\eta_j$ is approximatly zero for some frames. We write it as
\begin{align} \label{eq:test_error}
    E^{\textrm{test}}_k(t) = \frac{\|\eta_k(t) - \eta^{\textrm{model}}_k(t)\|_2}{A_k\: \sqrt{l}},
\end{align}
where $A_k$ is the nondimensional amplitude of the test soliton $k$, and $l$ is the number of spatial points. We also define the cumulative average error as
\begin{align} \label{eq:cumulative_error}
    E^{\textrm{cum}}_k(t) = \frac{1}{t}\int_0^t E^{\textrm{test}}_k(\tau)\: \mathrm{d}\tau.
\end{align}

In the first row of \cref{fig:Error_modeling}, we show the error defined in \cref{eq:test_error} at each time step for both methods. Models built from WSINDy coefficients are less accurate than those built from Fourier coefficients for times between approximately $15$ and $25$. The WSINDy models reach a maximum mean error of 16.37\% of the amplitude, while the Fourier multiplier method stays below 8\% (maximum is 7.63\%). The second row in \cref{fig:Error_modeling} shows the cumulative average error. The cumulative average error in the last time step is $6.27\%$ for WSINDy, and $4.02\%$ for the Fourier multiplier method. 

\begin{figure}[H]
    \centering
    \includegraphics[width=.5 \textwidth]{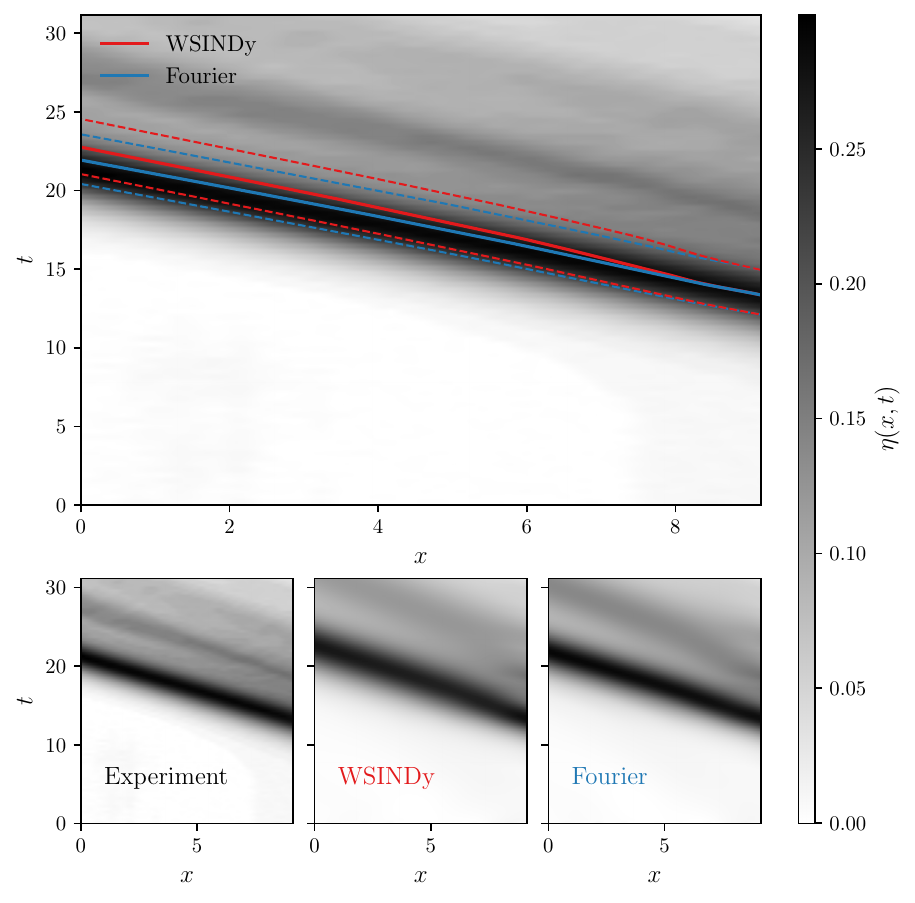}
    \caption{Heatmap of the experiment and the WSINDy and Fourier multiplier models. The top image shows the experiment as a heatmap. The solid lines show the crest (phase velocity), and the dashed lines indicate where the models have a surface elevation of 0.2 to illustrate dispersion. In the lower row, we show the full simulations and the experiment. This is for test soliton no. 3, for which both methods performed worst, based on cumulative average error.}
    \label{fig:Forward_time}
\end{figure}

In \cref{fig:Forward_time}, we plot the test case with the greatest cumulative average error in the final time step. The errors primarily manifest because of slight mispredictions of phase speed and dispersion, causing the predicted soliton to gradually move out of phase with the observed one. The coefficients obtained by the Fourier method yields phase speed and dispersion closer to the actual values, and errors do not grow in time as fast as for the WSINDy. The simulations for the training solitons give the same type of error as the testing, as shown in the appendix.  

\section{Methods comparison and discussion} \label{sec:discussion}

The equation-discovery method we present here is different from previous methods based on using discrete Fourier transforms for this purpose; these either use the Fourier transform to calculate derivatives in real space \cite{reinbold2019data,liu2021automated,patel2021physics} as in \cref{eq:Fourier_time_derivative}, denoise by removing high frequencies \cite{zhang2021robust,cao2023machine} like the low-pass filter applied here (see \cref{fig:Spectral_projection}), or Fourier transform all available variables \cite{tang2023fourier, cao2020machine}. In \citet{tang2023fourier, cao2020machine}, for instance, the authors find coefficients in $(\xi,\omega)$ space (spatial and temporally transformed spaces), whereas our method of finding coefficients in the $(\xi, t)$ (reciprocal-position, real-time) space has not been proposed before. Since the Fourier transform is unitary, transforming in time should not change the minimization. However, it is often not workable in practice to crop the data in time as we did in the transformed spatial variable. Neither is it advisable to crop the data before transforming because 
the reduced number of points would lead to an unworkably low resolution in frequency space. Cropping the data is a necessary step in our application because including time steps where the solitons leave or enter the frame leads to unusable results. 
Therefore, we believe that our procedure represents a significant contribution in applying these methods to real world data.

The primary similarity between the methods is that they avoid calculating high-order derivatives on imperfect data, which would greatly amplify noise. Rather, both our methods remove high-frequency signal via different types of windowed averaging. In WSINDy, derivatives are approximated by local averages weighted with analytically differentiated test functions with compact support, whereas in the Fourier method, we minimized the noise in two steps by cropping the data: similar to choosing the domain size for the test functions in WSINDy, we chose two frames when the soliton was centered in the frame to minimize the discontinuity at the boundaries caused by the limited window size. In the second step, the spatial Fourier transform was applied to the data, shifting discontinuities and noise to higher frequencies. The highest frequencies were discarded, and the Fourier multipliers were only found for the lowest wave numbers. To further reduce noise artifacts in WSINDy, the integrations were repeated at $K$ random locations in the data, yielding a mean over different spatio-temporal subdomains of $H(x,t)$, thereby making the WSINDy regression nonlocal. Thus, the $K$ integration domains are a direct parallel to the truncation in Fourier space.

One of the main differences between the Fourier multiplier method and the WSINDy approach is in the step at which they assume the types of terms present in the PDEs. In WSINDy, a specific choice must be made regarding the terms included in the library \textit{before} constructing the matrices in weak space and formulating the minimization problem. Although the sparse regression technique is designed to eliminate terms from the differential equation that do not contribute to the dynamics, it is essential to include the correct terms from the outset. In contrast, the Fourier multiplier method allows the operators $\ell$ and $n$ to be considered exact. The least squares minimizer identifies these operators without any prior assumptions about the type of equation. 
One may subsequently choose to restrict the operators to certain forms, for example, to facilitate physical interpretation as we do in this work, but it is not necessary.
The polynomial fit could have been achieved using sparse regression, which would allow the regression to estimate both coefficient magnitudes and the polynomial order. Although this would have provided another parallel between the approaches, another least squares was chosen at this step to maintain simplicity in our discovery process. The way we implement it here, the Fourier multiplier method demonstrated that we could obtain good results with a straightforward approach, given some prior knowledge of the system, e.g., by keeping only the odd parts of the multipliers. 

Another difference is the computational time of each approach. The implementation of the WSINDy algorithm has several performance disadvantages. As noted, for instance, by \citet{wareing2025data}, using PySINDy with a large data matrix or dataset can lead to significant memory issues. The memory requirements depend on the number of terms in the library and the value of $K$, which can quickly exceed the capacity of a standard desktop computer even if the data set itself is not large. Additionally, the time required to construct the total library is proportional to $K$, $N_x \times N_t$, and the number of terms. While WSINDy offers a statistical overview of potential equations, the process becomes time-consuming when $M$ is large, even when executed in parallel. In the Fourier multiplier method, the time derivative and the fast Fourier transform are calculated once. The regression is then always applied to the same operators $\ell$ and $n$. Therefore, in terms of the time required to obtain an equation, the Fourier method is superior.

Finding a regression that minimizes error in weak or Fourier space does not guarantee accurate modeling outcomes. Minimizing only the regression error could lead to solutions that are unstable with respect to small changes in the coefficients. It can lead to PDEs that are unstable with respect to the data, but whose modeling outcomes are not. The models are built by solving the PDEs as time-dependent boundary-value problems. When compared with the experimental data, we found that the mean modeling errors were below 20\% of the amplitude at all time steps. The errors for the test solitons were of the same order as those for the training solitons (see appendix). Since the equations were not discovered with an imposed criterion for producing good simulations, the comparable errors suggest that we have not overfitted the equations to the training solitons. Instead, we have found an equation that describes all solitons we created in the flume on the day of acquisition equally well. Reviewing the simulation, however, reveals two main inaccuracies: the phase speed is slightly off, and the dispersion is greater than in the experimental data. These discrepancies are more pronounced in the WSINDy models, suggesting that the balance among the terms is not optimal. The coefficient accompanying $\partial_x H$ sets the phase velocity and is underestimated by WSINDy compared to the Fourier method, as clearly seen by the colored lines in \cref{fig:Forward_time}. In the KdV equation, the steepening $\partial_x H^2$ balances the dispersion $\partial^3_x H$, allowing sech$^2$ shaped solitons to maintain their shape. The two methods found different balances, which is why they disperse differently as well. The Fourier multiplier balance is sharper, allowing solitons to maintain their shape for longer and disperse less. However, both make reasonable models for the time the waves are in the frame, especially given that \cref{fig:Forward_time} shows the worst-performing models.

\section{Conclusions}

In conclusion, we have applied two data-driven methods to find an equation from simple experimental data on shallow-water solitons. With the first method, we adapted an existing procedure, WSINDy \cite{brunton2016discovering}. The second relies on Fourier multipliers and is novel. Both methods make very few \emph{a priori} assumptions about the type of terms the discovered PDE may contain. Although the two methods are very different in their approach, they extract the exact same PDE from our data with only slightly different coefficients. It shows that even with our makeshift wave flume and 'GoPro physics' approach, the data inherently contains the governing equation which is robust against the choice of method to extract it. 

With our experimental data and two separate methods, we are able to validate and test both the new Fourier method and WSINDy and compare the methods to each other. By considering withheld laboratory cases that were not used for equation discovery and comparing with prediction from solving the discovered PDEs forward in time with matching initial conditions, allows a direct test of the predictive skill. The mean error of both PDEs remains below 20\% of the amplitude, while there is some systematic mismatch due to slightly wrong phase velocity and overprediction of dispersion. 

In terms of computing time, the Fourier method is clearly superior to WSINDy. In addition to constructing a library of candidate equation terms, WSINDy involves performing $M$ repetitions, each involving $K$ integrals, with $M$ and $K$ in the order of $1000$-$5000$ for final results, quickly becoming unwieldy for a standard desktop computer. In contrast, the Fourier method is (given the choices in \cref{subsec:results_FT}) deterministic and can be completed with such a machine in a few seconds with limited memory use. The predictions of the Fourier method are also slightly more accurate than those of WSINDy for the data set we have considered. 

We have demonstrated that although our experiment is extremely simple with no calibration or tuning, the underlying equation describing shallow-water solitons is inherently present in the data and can be extracted regardless of the method used. Key to our approach was to assume as little as possible \textit{a priori} about the form of the underlying equation, and to identify one that is easy to understand. It is worth noting that the equation we find is not exactly the Korteweg-de Vries equation from idealised theory, but a generalised version including a significant fifth-order term. Our goal here is not a deep understanding of the particularities of the flow in our makeshift lab, yet one could easily imagine using such a simple set-up as a low-cost precursor to a full-scale experiment. Discovering interesting aspects such as a significant fifth-order derivative contribution could point to where new aspects are to be found.

\subsection*{Acknowledgements}

The research was co-funded by the Research Council of Norway (\emph{iMOD}, grant 325114) and the European Union (ERC CoG, \emph{WaTurSheD}, grant 101045299).
Views and opinions expressed are, however, those of the authors only and do not necessarily reflect those of the European Union or the European Research Council. Neither the European Union nor the granting authority can be held responsible for them.
We have benefited from discussions with Professor J. Nathan Kutz. 

\appendix
\section{Experimental details} \label{sec:experments}
\begin{figure}[htb]
\centering
    \includegraphics[width=0.5\textwidth]{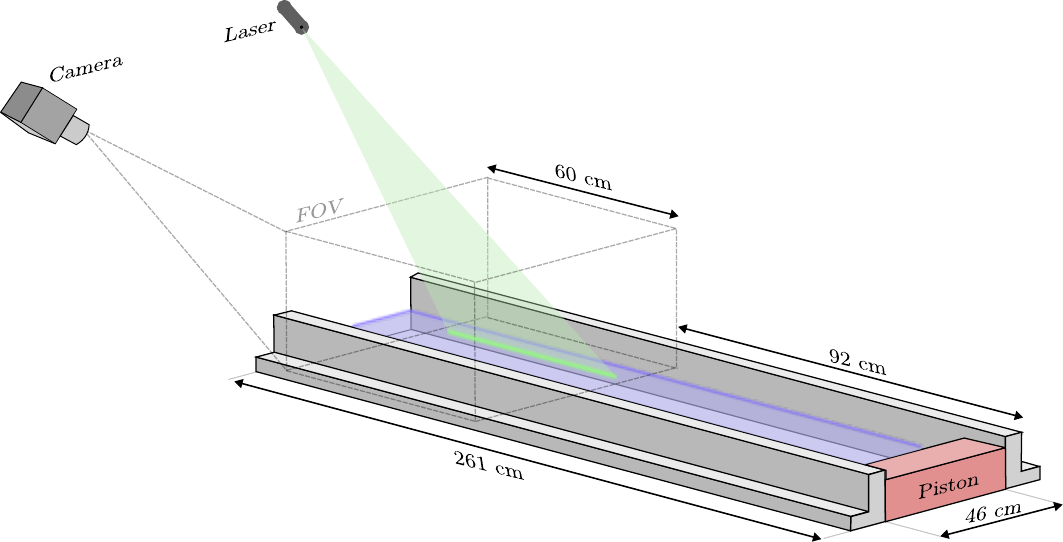}
    \caption{The dimensions of the setup are shown as well as the field of view. The laser was pointed from upstream, and the induced fluorescence was captured with a camera from the side. The field of view of the images was about 60 cm. The water depth was $h\approx32$mm.}
    \label{fig:experimental_setup}
\end{figure}

The experiment was conducted in the Strømningsteknisk lab at NTNU. Solitons were generated in a channel by manually pushing a piston, making it hard to control parameters such as wave speed and height. The surface height was recorded using planar laser-induced fluorescence (LIF). The details of the experimental setup and its dimensions are summarized in Figures \cref{fig:experimental_setup} and \cref{tab:experimental_setup}. In Section \cref{sec:SM_data}, we briefly discuss the train/test split and the data itself.

\subsection{Equipment}
The equipment used in this study is detailed in Table \cref{tab:experimental_setup}, and the experimental setup is illustrated in Figure \cref{fig:experimental_setup}. It is worth noting that, although we used a Photron mini WX100 camera, we chose a resolution and frame rate comparable to those found in most modern smartphones. We used a frame rate of 50 frames per second and a shutter speed of $1/200$s. Thus, this experiment can be easily reproduced with minimal equipment: just a modern smartphone and a kind of water channel. The Laser-Induced Fluorescence (LIF) measurements enhanced the contrast of the water surface against the background. There are also methods available to identify the surface using ordinary videos; for more information, see, for instance, \citet{cao2024identification}.

\begin{table}[htb]
    \centering
    \begin{tabular}{ll}
    \hline\hline
        \textbf{Camera} & Photron mini WX100 \\
        \textbf{Camera filter} & 510 nm low pass\\
        \textbf{Camera lense} &  50 mm\\
        \textbf{Laser} & Blue laser pointer\\
        \textbf{Sheet optics} & f-20mm\\
        \textbf{Fluorescein} & 100mg\\
        {\textbf{Water tank size}} & 2 m$^3$\\
    \hline\hline
    \end{tabular}
    \caption{Equipment that was used in the experiment. The Fluorescein and water tank size values are approximate.}
        \label{tab:experimental_setup}
\end{table}

The setup depicted in \cref{fig:experimental_setup} was positioned on top of a table within a larger water tank. Therefore, the volume of water inferred by the dimensions in the figure does not match the \textit{Water tank size} given in \cref{tab:experimental_setup}. 

\subsection{Data} \label{sec:SM_data}
\begin{figure}[htb]
    \centering
    \includegraphics[width=0.5\textwidth]{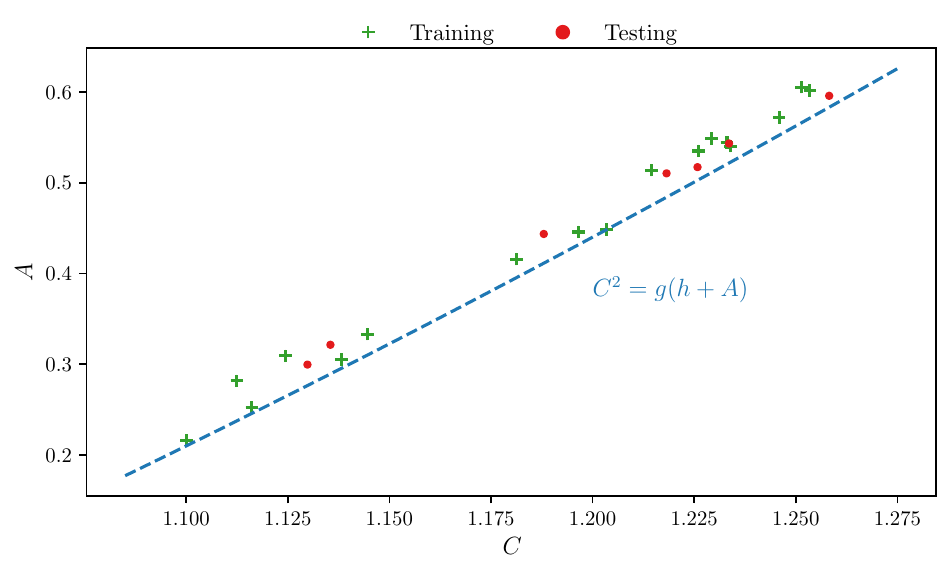}
    \caption{The amplitude and phase velocity of the solitons used in training and testing. The dashed line is the theoretical phase velocity for gravity-surface waves \cite{mei2005theory}.}
    \label{fig:velocity_amplitude}
\end{figure}

The collected data contains 25 solitons, each with varying amplitudes and wave speeds. This data was divided into two sets: 18 solitons for training and 7 solitons for testing. The selection of solitons for each set was not biased towards achieving lower errors; we used the first 18 solitons filmed for training and the last 7 for testing. In \cref{fig:velocity_amplitude}, we have plotted the dimensionless amplitude and phase velocity of the solitons.

\section{Edge detection}
The primary method for edge detection utilized in this process is the Canny algorithm \cite{canny1986computational}. The overall procedure can be divided into several sequential steps. Please note that the description provided is for a single video frame, and these steps are repeated for all frames throughout the video. We do not make any assumptions about the relationship between frames, meaning that edge detection for each frame is conducted independently. The steps are outlined below:\\

\begin{enumerate}
    \item Apply the Canny edge detection algorithm, which consists of four stages: (i) gradient calculation, (ii) non-maximum suppression, (iii) double thresholding, and (iv) edge tracking by hysteresis. In stage (iii), we set a minimum threshold of 10 and a maximum of 100. 
    \item Identify the uppermost edge and discard the rest.
    \item Convert the frame into two position vectors, $x$ and $\eta(x)$, and apply smoothing to the latter using the code provided in \cref{code:CMG_smoothing}.
\end{enumerate}

The implementation of the described steps is presented in \cref{code:Edge_detection}. This code takes a frame (represented as an n-dimensional array) as input and returns the position vector $x$ and the surface elevation $\eta(x)$ in pixel units. In the processing of all frames, cropping is done uniformly, which means that the value of $x$ remains consistent across all frames and videos. The final step in the preprocessing, although not shown in the listings below, involved converting $x$ and $\eta(x)$ to meters and then subtracting the mean water depth, $h$.

\begin{tiny}
\begin{lstlisting}[language=Python, caption={The edge detection preformed in Python, done for each frame independently.}, label={code:Edge_detection}]
def __get_edge__(frame: ndarray) -> (ndarray, ndarray):
    # Step 1 and 2, convert to greyscale and identify edges with Canny
    frame_grey = rgb_to_gray(frame)
    edges = Canny(frame_grey, thresh_min=10, thresh_max=100)

    # Step 3, get the uppermost edge
    Upper_edge = zeros_like(edges)
    for i in range(len(edges.T)):
        first_nonzero_index = argmax(edges[:, i])
        Upper_edge[:, i][first_nonzero_index] = 1

    x_coords = arange(Upper_edge.shape[1])
    y_coords = argmax(Upper_edge, axis=0)

    # Step 4, smooth in Matlab
    y_smooth = CMG_smoothing(y_coords)

    return x_coords, y_smooth
\end{lstlisting}
\end{tiny}

\begin{tiny}
\begin{lstlisting}[language=Matlab, caption={The smoothing function in \cref{code:Edge_detection}, inspired by ref. \cite{cao2024identification}.}, label={code:CMG_smoothing}]
function y_smoothed = CMG_smoothing(y)
    nn = 121;
    first = y(1);
    last = y(end);

    y_extended = [repmat(y(1), 1, floor(nn/2)), y, repmat(y(end), 1, floor(nn/2))];

    gaussian = smoothdata(y_extended,'gaussian',nn,'includenan');
    dummy = smoothdata(y_extended,'sgolay',nn);
    gaussian(1:round(nn/10)) = dummy(1:round(nn/10));
    gaussian( round(1:round(nn/5+(3/4)*(nn/5)) )) = smooth( gaussian( 1:round(nn/5+(3/4)*(nn/5)) ));

    y_extended_smoothed = gaussian-0.5;
    y_smoothed = y_extended_smoothed(floor(nn / 2) + 1 : length(y_extended_smoothed) - floor(nn / 2));
end
\end{lstlisting}
\end{tiny}

\section{Regression-implementation}
In this section, we explain the implementation of the Elastic net to the matrix equation $\bm{X} = \Theta \bm{c}$. The Elastic net is a regression method that solves a sparsity-promoting optimization problem. The loss function we are aiming to minimize is
\begin{align} 
    \hspace*{-2mm}\displaystyle \mathcal{L}(\bm{c}) = \frac{1}{2K} || \bm{X} - \Theta \bm{c}||^2_2 + \alpha \sigma ||\bm{c}||_1 + \frac{\alpha}{2} (1-\sigma)||\bm{c}||_2^2.
\end{align}
The $\alpha$ variable determines the sparsity of the solution $\bm{c}$, and $\sigma$ decides the relative weighting of the $L_1$ and $L_2$ norms. We find the optimal $\alpha$ and $\sigma$ values before estimating the sparse coefficients. This is done with the function \verb|sklearn.linear_model.ElasticNetCV| provided by scikit learn \cite{scikit-learn}. It compares ten model discoveries of a 10\%-90\% test and train split for the cross-validation, which has been demonstrated to provide test error rate estimates with low bias and variance \cite{james2013introduction}. The algorithm used in the main text is presented below. Simply put, it finds the optimal $\alpha$ and $\sigma$ values, removes features until four or fewer are left, and then estimates the actual coefficients using least squares. 
\begingroup
\renewcommand{\figurename}{Algorithm}
\begin{figure}[H]               
\caption{\texttt{STElasticNetCV}($\Theta$, $\bm{X}$, $\tau$, iters)}
\label{alg:stelasticnetcv}
\begin{algorithmic}[1]
  \Require $\Theta \in \mathbb{R}^{K \times n}$, $\bm{X} \in \mathbb{R}^{K}$, tolerance $\tau$, iters
  \State Define hyperparameter values:
  \Statex $\alpha \gets \texttt{logspace}(-12, 1, 100)$
  \Statex $\sigma \gets \texttt{linspace}(0.1, 1.0, 10)$
  \State Train ElasticNet using 10-fold cross-validation:
  \Statex $\bm{c} \gets \text{\texttt{ElasticNetCV}}(\Theta, \bm{X}, \alpha, \sigma, \text{cv}=10)$
  \Statex $\bm{c}_{\text{temp}} \gets \bm{c}$
  \State \textbf{Threshold small coefficients}:
  \Statex $\mathcal{S} \gets \{j : |c_j| < \tau \}$; \ \ $\bm{c}_{\mathcal{S}} \gets 0$
  \If{$\lVert\bm{c}\rVert_0 > 4$}
    \If{$\bm{c} = \bm{c}_{\text{temp}}$} \State $\tau \gets \tfrac{3}{2}\tau$ \EndIf
    \State $\bm{c}_{\mathcal{S}^c} \gets \texttt{STElasticNetCV}(\Theta_{:,\mathcal{S}^c}, \bm{X}, \tau, \text{iters}-1)$
  \EndIf
  \State Identify active feature indices: \ $\mathcal{A} \gets \{j : c_j \neq 0\}$
  \State Final OLS: \ $\bm{c}_{\mathcal{A}} \gets \Theta_{:, \mathcal{A}}^{-1}\bm{X}$
  \State \textbf{Return} $\bm{c}$
\end{algorithmic}
\end{figure}

\section{Amplitude and cumulative error}
\label{app:Amplitude_error}

In \cref{fig:error_amplitude}, we have plotted the end cumulative error $E^{\text{cum}}_k(T)$ against the nondimensional amplitude of the waves $A_k$ in the training and test set. The Fourier multiplier equation consistently produces lower modeling errors than the WSINDy one. Note that we have not used the 'forward solver' in the equation discovery for either the training or test set. In \cref{fig:error_amplitude}, we show the cumulative error for both sets to show that test solitons yield comparable modeling errors as the solitons that were trained on. 

\begin{figure}[htb]
    \centering
    \includegraphics[width=0.5\textwidth]{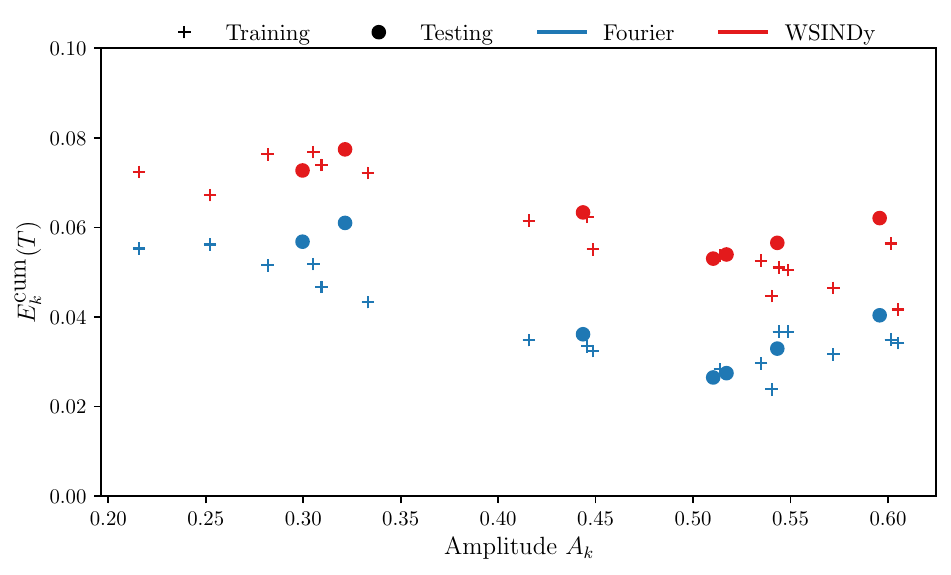}
    \caption{The cumulative errors sorted by amplitudes for both the training test solitons with the WSINDy and Fourier multiplier coefficients. The amplitude is nondimensional.}
    \label{fig:error_amplitude}
\end{figure}

\bibliographystyle{apsrev4-1}
\bibliography{refs}

\end{document}